\documentclass[reprint,amsmath,amssymb]{revtex4-2}

\usepackage{graphicx}
\usepackage[font=normalsize]{caption}
\usepackage{xcolor}

\begin{document}


\title{Non-local Andreev reflection through Andreev molecular states in graphene Josephson junctions}

\author{Eduárd Zsurka}
\email{eduard.zsurka@ttk.elte.hu}
\affiliation{Department of Physics of Complex Systems, Eötvös Loránd University, Budapest, Hungary}
\author{Noel Plaszkó}%
\affiliation{Department of Physics of Complex Systems, Eötvös Loránd University, Budapest, Hungary}
\author{Péter Rakyta}
\affiliation{Department of Physics of Complex Systems, Eötvös Loránd University, Budapest, Hungary}
\affiliation{Wigner Research Center for Physics, 29–33 Konkoly–Thege Miklos Str., H-1121 Budapest, Hungary}
\author{Andor Kormányos}
\email{andor.kormanyos@ttk.elte.hu}
\affiliation{Department of Physics of Complex Systems,
Eötvös Loránd University, Budapest, Hungary}

\begin{abstract}

We propose that a device composed of two vertically stacked monolayer graphene Josephson junctions  can be used for Cooper pair splitting. 
The hybridization of the Andreev bound states of the two Josephson junction can facilitate non-local transport in this normal-superconductor hybrid structure, 
which we study by calculating the non-local differential conductance.  Assuming that one of the  graphene layers is electron and the other is hole doped, 
we find that the  non-local Andreev reflection can dominate the differential conductance of the system. 
Our setup does not require the precise control of junction length, doping, or superconducting phase difference, which could be an important advantage 
for experimental realization. 

\end{abstract}

\maketitle


Quantum entangled particles have numerous potential applications in fields such as quantum communications or quantum cryptography. 
Thus, practical schemes of producing entangled particles are of fundamental interest~\cite{Vidal2003}. 
One of the most promising candidates for creating entangled electron states is based on  spin singlet Cooper pairs. 
It was proposed that if the electrons of a Cooper  pair can be extracted coherently and separated spatially,
they can  serve as a source of entangled electrons \cite{Recher,Lesovik_2001}.
This  process is  known as Cooper pair splitting  (CPS).
{As discussed in, e.g., Refs.~\cite{Samuelsson-CPS,Prada2004}, the key physical process to achieve CPS is the non-local or crossed Andreev reflection (CAR)}.

Although the first observations of Cooper pair splitting were made in metallic nanostructures \cite{beckmann-exp,russo-exp}, devices that use two quantum dots (QDs) have 
garnered the most attention in this field. The charging energy on the QDs prohibits the double occupancy on each dot, 
leading to the suppression of electron cotunneling (EC). EC is a competing process with CAR and it should be suppressed in order to achieve CPS. 
Experimentally CPS has been achieved in QD devices  realized in InAs and InSb nanowires~\cite{hofstetter1,hofstetter2,das,ueda,Kurtossy2022,Kouwenhoven-CAR-2022}, 
carbon nanotubes~\cite{hermann,schindele}, graphene based QDs~\cite{brange2021dynamic,tan, borzenets}, and recently in  2DEGs~\cite{Poschl}. 
Alongside the experimental effort, substantial theoretical work has also been devoted to the study of CPS in QD based 
devices~\cite{Recher,Falci-th,Lesovik_2001,Walldorf2020,Wimmer-CAR}. 

A different approach to suppress EC with respect to CAR  makes use of  features  in the density of states of semiconductors \cite{Cayssol,Veldhors-CPS}. 
Since this approach does not necessitate QDs, it  should make the fabrication of CPS devices simpler. 
Regarding monolayer graphene,   Ref.~\cite{Cayssol}
predicted that pure CAR could be achieved in a $n$-type graphene$-$superconductor$-p$-type graphene junction, if the doping of the graphene is smaller 
than the superconductor pair potential $\Delta_0$. In this case, the vanishing density of state of graphene at the Dirac point allows the elimination 
of processes that suppress CAR. However, due to the charge fluctuations around the Dirac point, which
are usually larger~\cite{Xue_2011,Mayorov2012} than the value of $\Delta_0$ of most superconductors, such a low doping  is  difficult to achieve experimentally.  
The problem of charge fluctuations can be mitigated, to 
some extent, by using bilayer graphene~\cite{Park-CAR}, because the larger density of states allows a better 
control of residual doping levels~\cite{Efetov2016,Park-CAR}. 
Recently, the signatures of CPS have  also been observed in multi-terminal ballistic graphene-superconductor structures~\cite{pandey2020ballistic}. 
{Another recent theoretical proposal~\cite{Soori-PRB2017,Nehra2019} suggested that the CAR probability can be enhanced in a device
where the central region consists of two, coupled one-dimensional superconductors and two normal leads are attached on each side to one of the  superconductors. 
The central region effectively constitutes  a superconducting QD. The CAR can be resonantly enhanced by tuning the superconducting phase difference between the 
one-dimensional superconductors to $\phi\approx \pi$ and then adjusting the chemical potential of the superconductors.}

\begin{figure*}
    \includegraphics[width=0.85\linewidth]{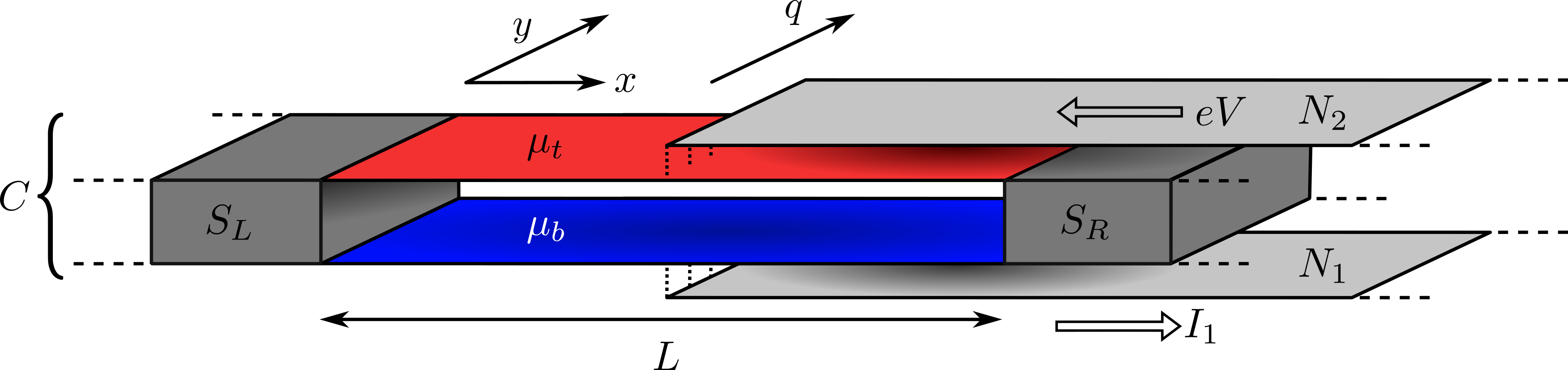}
    \caption{A schematic of the device. Two graphene monolayers (red and blue) of length $L$ and doping $\mu_t$ and $\mu_b$ are placed above each other and are connected at either side to superconducting leads $S_L$ and $S_R$ (dark gray). Normal leads $N_1$ and $N_2$ (light gray) are connected to each graphene layer at $x=L/2$. Translation invariance in the $y$ direction is assumed.
    {To study the properties of the device, we calculate the dependence of the current $I_1$ in $N_1$, when a voltage $V$ is applied to $N_2$}.}
    \label{fig:setup}
\end{figure*}

In this work we propose that  an approach based on Andreev molecular states~\cite{Girit-ABS-molec,Nazarov-ABS-molec} can also help to 
achieve CAR dominated transport. It was suggested that Andreev bound states (ABSs) in closely 
spaced Josephson junctions can overlap and hybridize forming Andreev molecular states (AMSs).  
We study the possibility of CPS in a setup that harbors AMSs. The device consists of two graphene 
Josephson junctions displaced vertically with respect to  each other, (see Fig.~\ref{fig:setup}) such that the ABSs in the two junctions can hybridize. 
This type of graphene JJ has  recently been studied experimentally in Ref.~\cite{Indolese2020}, 
focusing on superconducting interference device type operation and quantum Hall physics.
We calculate the non-local, non-equilibrium differential conductance through the device, when  two normal leads are weakly connected 
to the graphene layers, as shown in Fig.~\ref{fig:setup}. 
Our most important finding is that CAR can be larger than EC even if the doping of the graphene layers is significantly larger than the value of $\Delta_0$.  
Therefore, the CAR should be less affected by charge puddles, which are present in graphene at low doping. 

\section{The model}
\label{sec:model}

The schematics of the proposed four-terminal device is shown in Fig.~\ref{fig:setup}. Two graphene monolayers (red and blue) of length $L$ 
are placed above each other. They are separated by an insulator such as hBN or vacuum in the center of the device, i.e., for $0<x<L$, meaning that 
there is no direct electrical contact between these two layers vertically.  
Two superconducting leads, $S_L$ and $S_R$ (dark gray) are attached to the edges of the top and bottom graphene layers, at $x=0$ and $x=L$. 
In addition, two normal leads (light gray) $N_1$ and $N_2$ are weakly coupled to the middle ($x=L/2$) of the top and the bottom graphene layer, respectively.  
We note that a similar layout for a single graphene JJ junction was used in Ref.~\cite{Bretheau2017} to determine the energy spectrum of ABSs. 

{In our calculations the description of both the normal and the superconducting regions is based on the nearest-neighbor
tight-binding model of graphene with in-plane hopping amplitude $\gamma_0$. 
The top and bottom graphene layers and the superconducting leads constitute the central region of the device, described by the Hamiltonian}
\begin{equation}
H_C=\begin{pmatrix}
      H_{gr}-\mu_t & 0 & W_{NS}\\
      0 & H_{gr}-\mu_b & W_{NS}\\
      W_{NS}^\dagger & W_{NS}^\dagger  & H_S - \mu_S
     \end{pmatrix}.
\label{eq:H_C}     
\end{equation}
{Here  $H_{gr}$ is the Hamiltonian of undoped monolayer graphene,  }
$H_S=\left(
\begin{array}{cc}
 H_{S_L} & 0\\
 0 &  H_{S_R}
\end{array}
\right)
$ 
{is the Hamiltonian of the superconducting leads  in the non-superconducting state.
The  leads $S_L$ and $S_R$ are modeled with Bernal stacked  multilayer graphene, with out-of plane hopping amplitude $\gamma_1$.
We assume that the top and bottom graphene layers  are perfectly aligned and denote the doping by $\mu_t$ [$\mu_b$] in the top [bottom] layer, 
while $\mu_S$ is the doping in $S_L$ and $S_R$. 
$W_{NS}$ describes the coupling between the graphene layers  and the superconducting leads with hopping amplitude 
$\gamma_{NS}=\gamma_0$, corresponding to a perfectly transparent interface. }

{Before superconductivity is introduced, the total Hamiltonian of the system  reads }
\begin{equation}
H_{tot}= \begin{pmatrix}
    H_C & W_1 & W_2\\
    W_1^\dagger & H_1 & 0\\
    W_2^\dagger & 0 & H_2
    \end{pmatrix}, 
\label{eq:H_tot}    
\end{equation}
{where $H_l=H_{gr}-\mu_l$ is the Hamiltonian of the normal leads $N_l$, with $l=1,2$.
The leads  $N_l$ are also modelled by monolayer graphene and their  doping  is kept fixed at  $\mu_l=0.1$\,eV.
We checked that the results discussed below do not strongly depend on $\mu_l$. 
$W_l$ describes the coupling between $N_l$ and the corresponding graphene layer (see Fig.~\ref{fig:setup}).}

\begin{figure*}[t]
    \centering
    \includegraphics[width=0.7\linewidth]{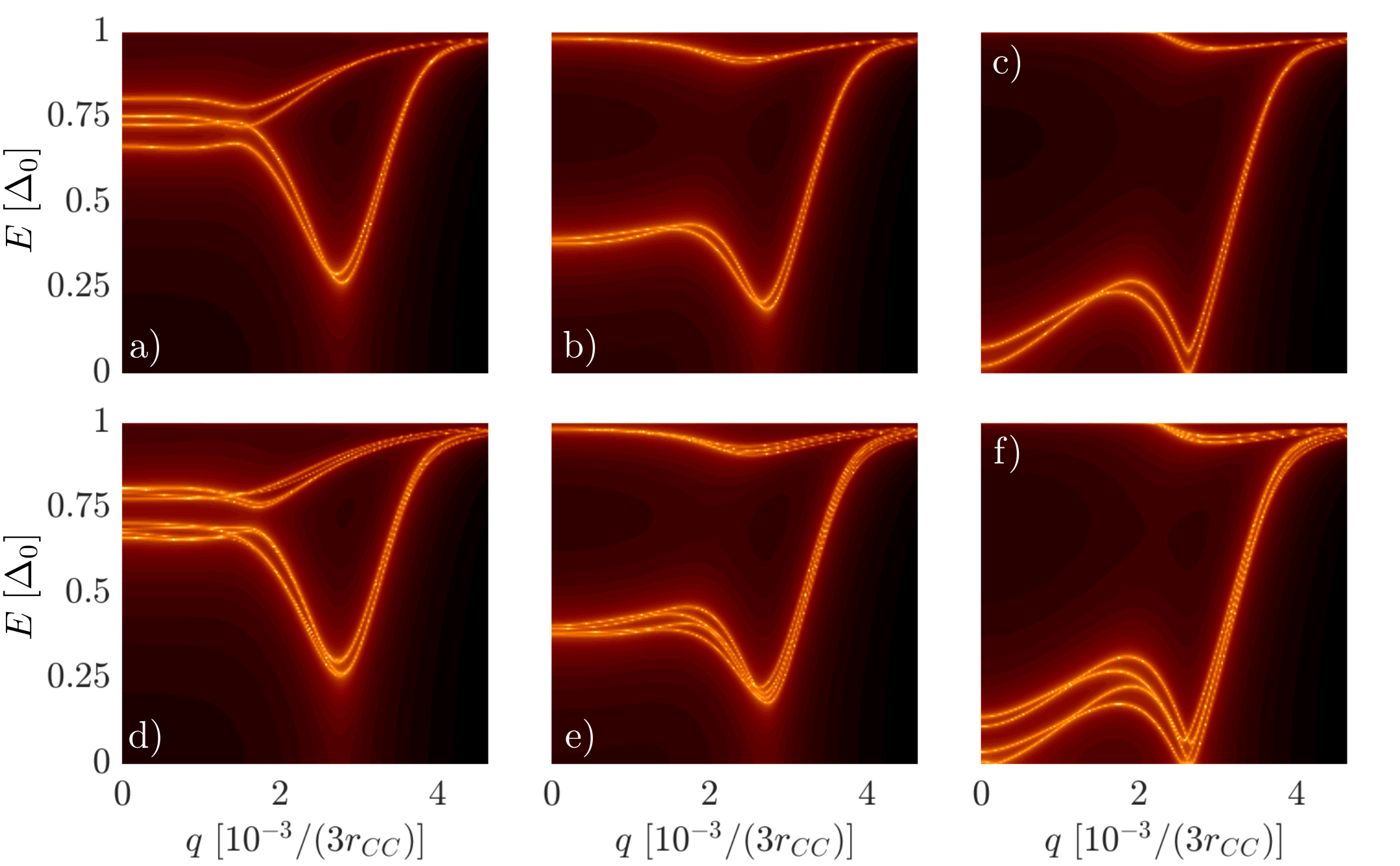}
    \caption{Local density of state calculations of Andreev bound states as a function of the momentum $q$, for three different superconducting phase differences $\varphi$.   
    a)-c) LDOS of ABSs in a monolayer graphene JJ, for $\varphi=0,\,\pi/2$ and $\pi$ going from left to right. d)-f) LDOS of AMS in the top graphene layer, i.e., when both graphene layers are connected to the superconductors, for $\varphi=0,\,\pi/2$ and $\pi$ going from left to right. We used doping $\mu_b=-\mu_t=5\Delta_0$ and  $E_{Th}=\Delta_0$ in all cases.}  
    \label{fig:DOS-1L-2L}
\end{figure*}

{To describe the transport  properties of this system when the leads $S_L$  and $S_R$ are superconducting, we used the approach based on
the Bogoliubov–de Gennes Hamiltonian. This can be compactly written as }
\begin{equation}
    \begin{pmatrix}
    H_{tot}-E_F & \tilde{\Delta}(x) \\
    \tilde{\Delta}^{*}(x) & -H_{tot}+E_F
    \end{pmatrix}
    \left(\begin{array}{c}
           \Psi_e\\
           \Psi_h
          \end{array}\right)
          =\varepsilon
          \left(\begin{array}{c}
           \Psi_e\\
           \Psi_h
          \end{array}\right),
\end{equation}
{where  $E_F$ is the Fermi energy, $\varepsilon>0$ is the excitation energy, $\Psi_e$ and $\Psi_h$ are electron and hole wave functions, respectively.
$\tilde{\Delta}(x)$ is a matrix which only has non-zero elements between degrees of freedom that belong to either $S_L$ or $S_R$. 
To describe superconductivity, an $s$-wave pairing potential is used, which is nonzero only in the superconducting 
 leads and changes in a  step-function manner at the normal-superconducting interface: 
 $\Delta(x)=\Delta_0[\theta(-x)+\theta(x-L)\mathrm{exp}({i\varphi})]$, where $\theta$ is the Heaviside function, 
 and $\varphi$ is the superconducting phase difference between $S_L$ and $S_R$. 
 The step-function change of the pair-potential at the boundary  is valid  if $\lambda_F^{(S)}\ll\lambda_F^{(t,b)}, \xi_0$~\cite{Titov2006}. 
 Here $\lambda_F^{(S)}$ and $\lambda_F^{(t,b)}$ 
 are the Fermi wavelength in the superconducting leads and central graphene layers and 
 $\xi_0=\hbar v_F/\Delta_0$ is the (in-plane)  ballistic superconducting coherence length, {$v_F\approx 10^6$m/s being the Fermi velocity of monolayer graphene.}
  We use highly doped superconducting leads with $\mu_S=0.8$eV, therefore the above condition is satisfied in all our calculations.
Since the in-plane $\gamma_0$ and out-of plane $\gamma_1$ hopping amplitudes in Bernal stacked multilayer graphene are different, 
it is intuitive to define an effective superconducting coherence length  $\xi_{\perp}\neq\xi_0$ associated with the out-of-plane hopping 
in the superconducting leads. One can expect that interlayer Andreev reflection from the top to the bottom graphene layers is only significant 
 if $d\lesssim \xi_{\perp}$, where $d$ is the vertical distance between these layers. 
 We explain how $\xi_{\perp}$ is estimated in Supplementary Information (SI), here we only mention than in all subsequent 
 calculations $d\ll \xi_\perp$. }

In the transport calculations we assume that a voltage $V$ is applied (with respect to $E_F$) to the top normal lead $N_2$ and the 
current $I_1$ is measured in the bottom normal lead $N_1$, as shown in Fig. \ref{fig:setup}.
{We calculate the non-local differential conductance $G(eV)=d I_1/d V$ which depends on CAR.
We are primarily interested in the case of wide graphene layers, where exact termination of the edges  does not matter
because the transport properties are determined by bulk states. Using hard wall boundary conditions~\cite{PhysRevLett.96.246802,PhysRevB.74.041401}, the
transverse wavenumber $q$ parallel to the $y$ direction is a good quantum number, see the SI for further details.
The numerical calculations discussed below were performed using the tight-binding framework implemented in the EQuUs \cite{equus} package.}


\section{Andreev molecular states}
\label{sec:AMS-DOS}

The Andreev reflection of quasiparticles at the graphene-superconductor  interfaces leads to the formation of correlated electron-hole states 
known as Andreev bound states, \cite{Titov2006,Titov2007,Manjarres2014,BenShalom2016,Bretheau2017,Banszerus2020} with energies $E_{n}\le \Delta_0$. 
Their presence in the proximitized graphene layers means that an induced gap $\Delta_{ind}$ appears in the graphene layers, which is smaller than the pairing potential $\Delta_0$ of the superconductors. 
{If the superconducting phase difference $\varphi$ is fixed, in ballistic systems the magnitude of $\Delta_{ind}$ is
determined by the smaller of two energy scales, namely, the bulk gap $\Delta_0$ and the  Thouless energy $E_{Th}=\hbar v_F/L$.}

For $\varphi=0$, when  $E_{Th} \gg \Delta_0 $, i.e., in the short junction regime $\Delta_{ind}\approxeq\Delta_0$. In the opposite case $E_{Th} \ll \Delta_0 $, the dominant energy scale is $E_{Th}$; this is the long junction regime where $\Delta_{ind}$ is considerably smaller than $\Delta_0$. 
Note that the ratio of $E_{Th}$ and  $\Delta_0$ can also be expressed as $E_{Th}/\Delta_0=\xi_0/L$, so that the short junction regime corresponds $E_{Th}/\Delta_0\gg 1$. 
In this work, we study devices with Thouless energy between $0.4\Delta_0$ and $3\Delta_0$.
Junctions with $E_{Th}$ in this range correspond to the intermediate length regime, where analytic results valid in the short \cite{Titov2006} or long \cite{Titov2007} regime of a Josephson junction (JJ) consisting of a single graphene layer do not strictly  apply. Taking $\Delta_0=1$\,meV, the aforementioned $E_{Th}$ 
values correspond to $L$ between $210$ and $1580$\,nm.

One can expect that in the setup shown in Fig.~\ref{fig:setup},  the ABSs formed in the two graphene layers can hybridize, 
leading to the formation of Andreev molecular states (AMSs) \cite{Girit-ABS-molec,Nazarov-ABS-molec}. 
In order to see the effects of ABS hybridization, we start by considering the properties of ABSs formed in individual layers, i.e., when one of 
the graphene layers, e.g., the bottom one is  disconnected from the superconductors and only the top one is connected. 
We also disconnect the lead $N_2$  and calculate the Green's function of the resulting graphene JJ. 
The spectrum of the ABSs is determined  performing local density of states (LDOS) calculations for energies $0\leq E\leq\Delta_0$. 
The LDOS is calculated as 
{the sum of the LDOS of electron and hole type quasiparticles
$\rho(E,q)=\rho^{e}(E,q)+\rho^{h}(E,q)=-(1/\pi)\mathrm{Im}(\mathrm{G}^R)$, where $\mathrm{Im}(\mathrm{G}^R)$ is the imaginary part 
of the retarded Green's function.} 
The LDOS is evaluated on $\sim 10$ unit cells of the top layer around $x=L/2$.
In Figs.~\ref{fig:DOS-1L-2L}(a)-(c) we show results for 
superconducting phase differences $\varphi=0,\,\pi/2$ and $\pi$, using $\mu_t=-5$\,meV and $E_{Th}=\Delta_0$. 
One can clearly see the appearance of multiple 
ABSs. Above $E=0$ there is an energy range where no ABSs are present indicating the  induced gap $\Delta_{ind}$. One can observe that as $\varphi$ increases from $0$ to $\pi$, the induced gap $\Delta_{ind}$ decreases and 
at $\varphi=\pi$ the induced gap is closed. This can be  shown analytically in both the short~\cite{Titov2006} and long~\cite{Titov2007} 
junction regime and also agrees with the experimental results of Ref.~\cite{Bretheau2017}.

Turning now to the bilayer setup of Fig.~\ref{fig:setup}, the distance between the graphene layers is taken to be $d=3.3$\,nm in our calculations, while we found that $\xi_{\perp}\approx 38$\,nm (see SI). 
Since $d\ll \xi_{\perp}$, the coupling between the ABSs can lead to the formation of Andreev molecular states~\cite{Girit-ABS-molec,Nazarov-ABS-molec}. 
This is shown in  Figs.~\ref{fig:DOS-1L-2L}(d)-(f), where
one can see the LDOS $\rho_t(E,q)$ calculated in the top graphene layer. 
At this stage the normal leads $N_1$ and $N_2$ are not yet connected to the graphene layers.
For AMSs  with energies $E_{n}\lesssim\Delta_0$ the relatively weak hybridization leads to only  minor modifications of the LDOS, c.f. Figs.~\ref{fig:DOS-1L-2L}(a)-(c). 
However, for  $\varphi=\pi$ there are AMSs with energy $E_n\gtrsim 0$  which are more strongly modified by 
interlayer hybridization [Fig.~\ref{fig:DOS-1L-2L}(f)]. One can also see that, similarly to the case of ABSs [Figs.~\ref{fig:DOS-1L-2L}(a)-(c)],
the magnitude of $\Delta_{ind}$ in the presence of AMSs can  be tuned by changing $\varphi$ [Figs~\ref{fig:DOS-1L-2L}(d)-(f)].

\begin{figure}[hb!]
    \centering
\includegraphics[width=1.0\linewidth]{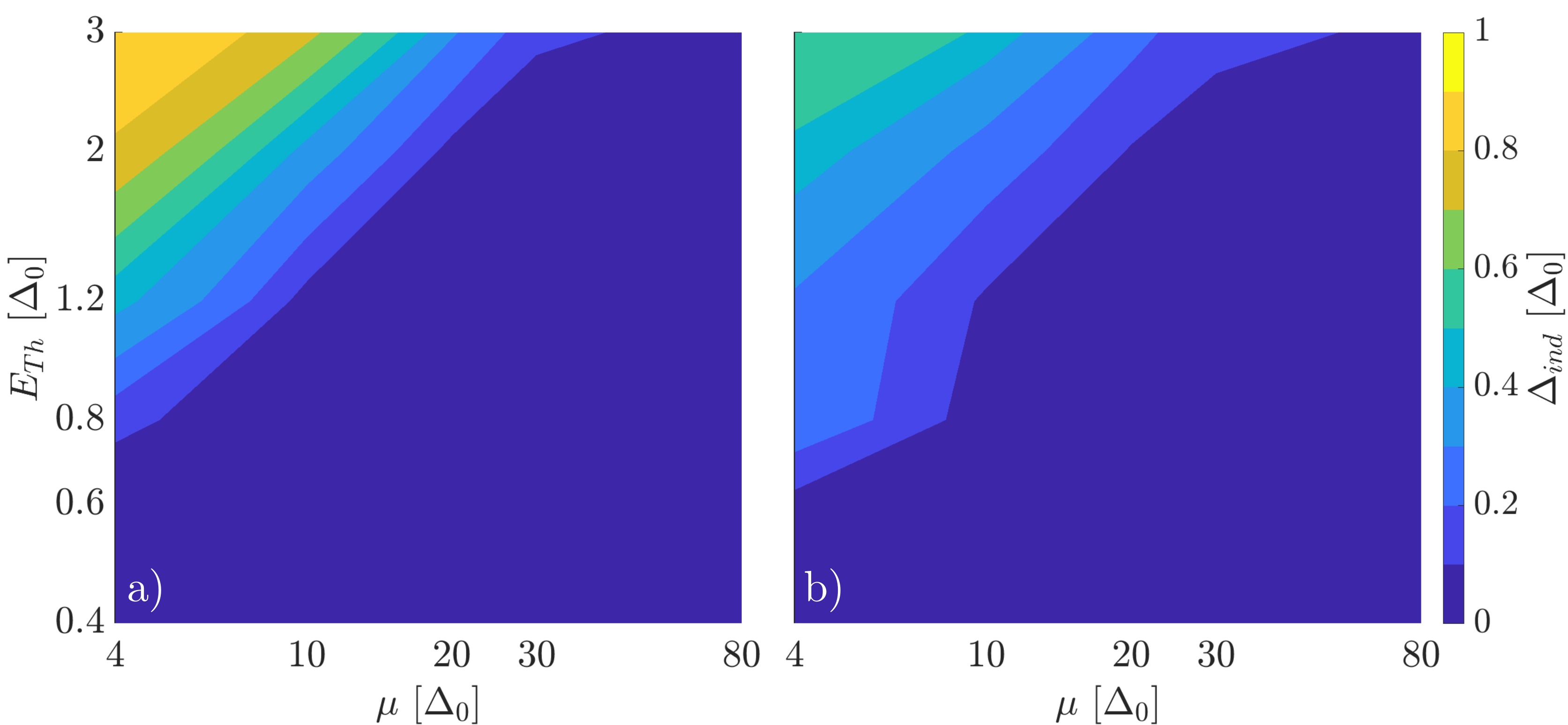}
    \caption{Size of the induced gap $\Delta_{ind}$ (in units of $\Delta_0$) as a function of the magnitude of the doping $\mu$, where $\mu=\mu_b=-\mu_t$,
    and Thouless energy $E_{Th}$, for superconducting phase difference a) $\varphi=0$ and b) $\varphi=\pi/2$.}
    \label{fig:DOSgap}
\end{figure}

One can expect that in order to have a finite interlayer transmission  of electrons from $N_1$ to $N_2$ 
in the bias window $|eV|\leq\Delta_0$, $\Delta_{ind}$ has to be smaller than $\Delta_0$. Therefore, $\Delta_{ind}$ is an important parameter of the device. 
We calculated $\Delta_{ind}$ as a function  of the doping $\mu$ and $E_{Th}$, where $\mu=\mu_b=-\mu_t$, see Fig.~\ref{fig:DOSgap}.
{The value of $\Delta_{ind}$ is extracted from LDOS calculations by determining the minimum of the AMS spectrum.}
We find that for {$\varphi=0$ and $E_{Th}=0.4\Delta_0,\,0.6\Delta_0$ the induced gap is suppressed $\Delta_{ind}\ll\Delta_0$ [Fig.~\ref{fig:DOSgap}(a)]. }
However, for larger values of $E_{Th}=0.8\Delta_0-3\Delta_0$ a
general observation is that $\Delta_{ind}$ is comparable to $\Delta_0$ for low doping, but increasing  $\mu$ leads to the reduction of $\Delta_{ind}$. 
For large enough doping the induced gap can be suppressed regardless of the Thouless energy for the $E_{Th}$ values we studied. In short, the condition $\Delta_{ind}<\Delta_0$ is satisfied
for a wide range of ($\mu$, $E_{Th}$) values. 
Note that tuning the doping changes not only $\Delta_{ind}$, but also the number of the AMSs.
Furthermore, as illustrated in Figs.~\ref{fig:DOS-1L-2L}(d)-(f), by increasing $\varphi$ the AMSs are shifted deeper into superconducting gap
{and $\Delta_{ind}$ decreases [Fig.~\ref{fig:DOSgap}(b)]}.
We find that in these ballistic devices for $\varphi=\pi$ the induced gap disappears regardless of the value of $E_{Th}$.


\section{Differential conductance}
\label{sec:diffcond}

We now discuss the transport through the central region of the device when  the normal leads $N_1$ and $N_2$ are attached, as shown in Fig.~\ref{fig:setup}.
We are interested in the dependence of $I_1$ in $N_1$ on the applied voltage $V$ to $N_2$.   
We restrict our study to voltages $|eV|\leq\Delta_0$, therefore one expects that the transport is mediated by the AMSs in the junction. 
We use the Keldysh non-equilibrium Green’s function technique~\cite{Cresti2003,Do2014,Pala2007,Bolech2005,Wu2004} to calculate $dI_1/d V=G(eV,q)$ 
for a given $q$ and then sum the contributions of the different $q$ values, see the SI for more details.
The differential conductance is given by 
\begin{eqnarray}
    G(eV,q) = -\frac{2e}{h}\mathrm{Re}\bigg\{\frac{d}{dV}\int dE \mathrm{Tr}\big[\tau_z W_{1}  \mathrm{G}^<_{C,1}(E,eV)\big]\bigg\},\nonumber\\
    \label{eq:dIdV2}
\end{eqnarray}
where $\tau_z$ is a Pauli matrix acting in the electron-hole space and $\mathrm{G}^<_{C,1}(E,eV)$ is the bottom lead--central region lesser Green’s function. 
To lighten the notations, the $q$ dependence of  $\mathrm{G}^<_{C,1}(E,eV)$ is not written explicitly. The differential conductance $G(eV)$ can be evaluated as
\begin{equation}
    G(eV) =\sum_q G(eV,q)=\dfrac{w}{2\pi} \int  G(eV,q)\,dq,
    \label{eq:G_total}
\end{equation}
where $w$ is the width of the junction in the $y$ direction. All calculations are performed at $T=0$ K temperature. 
 
{In order to obtain  an insight into the transport properties of this setup, let us first consider a simple model:
we assume that only a single AMS of energy $E_{AMS}$ is present, which extends over both graphene layers in the central region. 
{We neglect the $q$ dependence of the AMS  and assume  that  coupling between $N_1$ ($N_2$) and bottom (top) graphene layers is weak}.
According to the calculations detailed in the SI, the differential conductance can be approximated by
\begin{equation}
    G(eV)\approx \frac{4e}{h}\dfrac{(\Gamma_1^e - \Gamma_1^h )(\Gamma_2^e -\Gamma_2^h)  }{(eV-E_{AMS})^2+\Gamma^2},
    \label{eq:dIdV3}
\end{equation}
where $\Gamma_l^\alpha$  are level broadenings~\cite{Claughton1995} due to the coupling  of the electron [hole] ($\alpha=e[h]$) 
part of the AMS to the states in $N_l$ at energy $E_{AMS}$, and $\Gamma=\Gamma_1^e+\Gamma_1^h+\Gamma_2^e+\Gamma_2^h$.
Eq.~\eqref{eq:dIdV3} shows that the presence of an AMS results in a resonant peak of Lorentzian lineshape in the 
differential conductance, at $eV\approx E_{AMS}$.  
The signature of CAR dominated transport is $G(eV)<0$, meaning that an injected electron in $N_2$ is transmitted as a hole into $N_1$. 
The sign of $G(eV)$ is determined by the numerator in Eq.~(\ref{eq:dIdV3}), which depends on the difference 
between the level broadening of electron- and hole-like degrees of freedom of the AMS. 


{In the tunneling limit $\Gamma_l^{e(h)}$ depends on the product of the LDOS of the electron (hole) component of the AMS and
of the attached leads $N_l$. Since the leads are metalic, their LDOS is constant. Therefore
$\Gamma_l^e -\Gamma_l^h$ depends mainly on the difference of the LDOS of 
the electron and hole type quasiparticles in the AMS.  One can expect that this can be changed by two means:
firstly, by tuning the doping of the two graphene layers. Secondly,  since the 
AMS wave functions  depend on the superconducting phase difference $\varphi$, the LDOS can also be changed by tuning $\varphi$. 
Thus, this simple model suggests that one has two experimental knobs to tune the interlayer transmission and try to achieve CAR dominated transport.}

\begin{figure}[ht!]
     \includegraphics[width=0.96\linewidth,trim={0 0 3.2cm 0},clip]{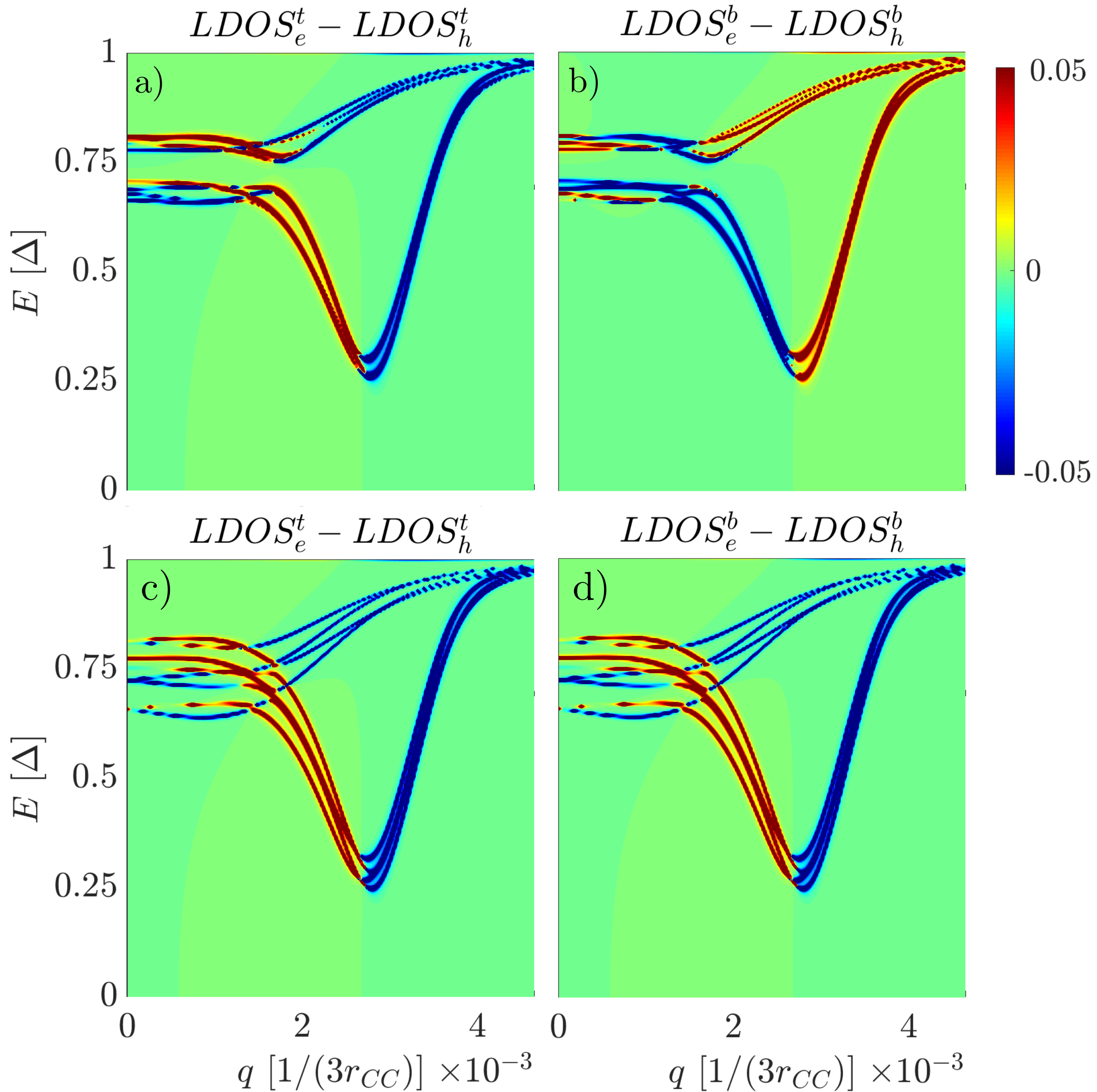}
     \caption{The LDOS difference of the electron and hole quasiparticles of AMSs in the top and bottom graphene layers. a) $\delta\rho_{t}(E,q)$ and b) $\delta\rho_{b}(E,q)$ for asymmetric doping and c) and d) for symmetric doping, respectively. Here $\varphi=0$, $E_{Th}=\Delta_0$ and $|\mu_b|=|\mu_t|=5\Delta_0$.}
     \label{fig:DOS_diff}
 \end{figure}
 
{As it can be seen in Fig.~\ref{fig:DOS-1L-2L}, for finite doping of the graphene layers, multiple AMS are present in our setup.
The result given in Eq.~(\ref{eq:dIdV3}) can be easily generalized  to this case (see the SI).  One finds that  $G(eV,q)$  defined 
in Eq.~(\ref{eq:dIdV2}) reads 
\begin{widetext}
\begin{equation}
    G(eV,q)= 
    \frac{4e^2}{h}\sum_{m, n}\dfrac{(\Gamma_{1,mn}^e(q) - \Gamma_{1,mn}^h(q) )(\Gamma_{2,mn}^e(q) -\Gamma_{2,mn}^h(q))  }{\left(eV-E_{m}(q)+i\Gamma_{mm}(q)\right)
    \left(eV-E_{n}(q)-i\Gamma_{nn}(q)\right)},
    \label{eq:G_mn}
\end{equation}
\end{widetext}
where the summation runs over the number of the AMSs, $\Gamma_{nm}$ depends on the product of the wave functions of the $n$th and $m$th AMS 
and 
$\Gamma_{nn}=\sum_{l,\alpha}\Gamma_{l,nn}^{\alpha}$.
The $m=n$ terms are Lorentzian resonances, this is the type of contribution  we have already discussed when we derived Eq.~(\ref{eq:dIdV3}).  
The  $m\neq n$ terms correspond to a ``cross-talk'' between different AMSs and they are affected by interference effects between different AMSs. 
Therefore, in general,  $G(eV,q)$ depends both on the LDOS and on the interference of the quasiparticle components of the AMSs. 
}

{Note, that in Refs.~\cite{Cayssol,Veldhors-CPS} the enhancement of  the probability of CAR is related to  the
DOS of the semiconducting leads,  which are attached to a central superconducting strip, and their different doping. 
In our setup  the leads $N_l$ are assumed to be metallic and their doping does not  play an important role. 
Moreover, as we discussed above,  in our case  quasiparticle interference also 
affects $G(eV)$, but as we will show in Sec.~\ref{sec:nonloc-G}
it does not lead to the type of resonant enhancement of CAR as in Refs.~\cite{Soori-PRB2017,Nehra2019}. 
These considerations  clearly show the difference between our proposal and those of Refs.~\cite{Cayssol,Veldhors-CPS,Soori-PRB2017,Nehra2019}.
}


%

\section{Negative non-local Andreev reflection}
\label{sec:nonloc-G}

\begin{figure*}[ht]
     \includegraphics[width=0.7\linewidth]{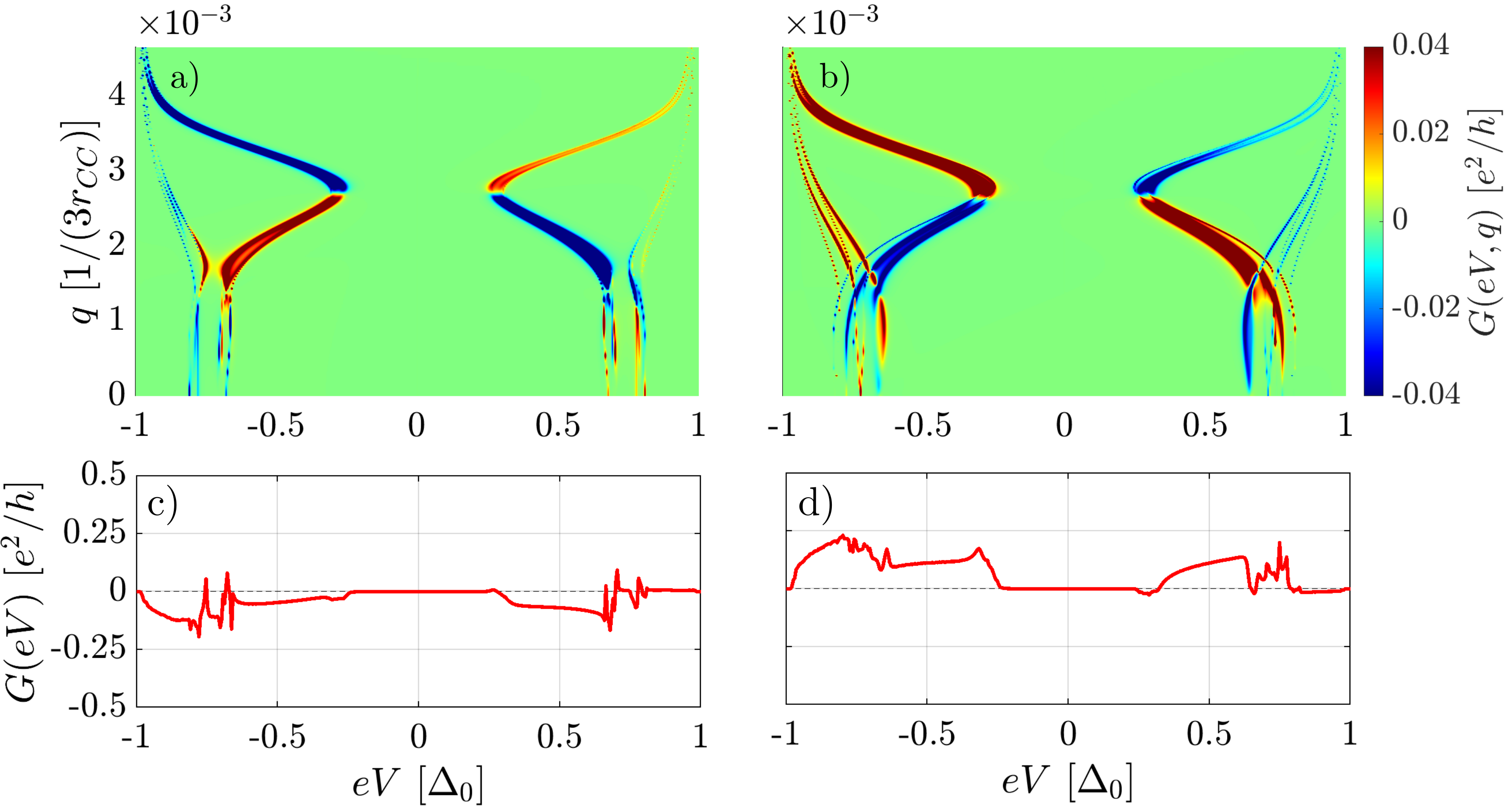}
     \caption{The $q$-resolved non-local differential conductance $G(eV,q)$ for a) asymmetric and b) symmetric doping for 
     the systems shown in Fig.~\ref{fig:DOS_diff}.
     c) and d) The total differential conductance $G(eV)$  corresponding to the case in a) and b), respectively.}
     \label{fig:Gevq}
 \end{figure*}

{We start with  calculations which illustrate the complex interplay of LDOS and interference related effects in the differential conductance.
In Figs.~\ref{fig:DOS_diff} we show the LDOS difference of the electron and hole  quasiparticles of AMSs 
$\delta\rho_{t}(E,q)=\rho^{e}_{t}(E,q)-\rho^{h}_{t}(E,q)$  ($\delta\rho_{b}(E,q)=\rho^{e}_{b}(E,q)-\rho^{h}_{b}(E,q)$) in the top (bottom) graphene 
layers. These results were obtained in the same way as the total LDOS   $\rho(E,q)$ in Fig.~\ref{fig:DOS-1L-2L}(d)-(f), 
i.e., evaluated on $\sim 10$ unit cells around $x=L/2$.
We consider two cases: $\mu_b=-\mu_t$ (asymmetric doping) and $\mu_b=\mu_t$ (symmetric doping) and the parameters of the calculations correspond to the 
case shown in Fig.~\ref{fig:DOS-1L-2L}(d).
In a given layer the sign of $\delta\rho^{}(E,q)$ depends on both the energy $E$ and the wavenumber $q$. However, one can clearly observe that 
for asymmetric doping $\delta\rho_{t}(E,q)$ has opposite sign to $\delta\rho_{b}(E,q)$. On the other hand, for symmetric doping  
$\delta\rho_{t}(E,q)=\delta\rho_{b}(E,q)$, which can be expected based on the inversion symmetry of the system. 
Since more than one AMSs gives contributions to  $\delta\rho^{}(E,q)$, these results cannot be directly related to individual broadening differences 
$\Gamma_{l,nn}^e - \Gamma_{l,nn}^h$, but they do illustrate the important effect of the doping of the two graphene layers. 
Furthermore, using the arguments put forward below Eq.~(\ref{eq:dIdV3}),
these results suggest that the sum of the $m=n$ terms in Eq.~(\ref{eq:G_mn}) gives a negative (positive) contribution 
to the differential conductance for asymmetric (symmetric) doping profile. 
}

{The contributions of the  $m\neq n$ terms in Eq.~(\ref{eq:G_mn}) is more difficult to visualize, but our  numerical
calculations indicate that they give an equally important contribution to $G(eV)$.  
To illustrate this point, in Figs.~\ref{fig:Gevq}(a) and (b) we show the $q$-resolved non-local differential conductance 
$G(eV,q)$ {for asymmetric and symmetric doping, respectively,} and weakly coupled normal leads $N_1$ and $N_2$. We used the same parameters as for the calculations  in Fig.~\ref{fig:DOS_diff}.
The non-zero matrix elements of $W_1$ and $W_2$ are on the order $0.1\gamma_1$, where $\gamma_1$ is the interlayer coupling in Bernal stacked graphene.
The general features in $G(eV,q)$ closely resemble the LDOS  in  
Fig.~\ref{fig:DOS_diff}, 
showing the important role of the AMSs in the non-local conductance for this relatively weak coupling between $N_1$, $N_2$ and the corresponding 
graphene layers. $G(eV,q)$ can be both positive and negative as a function of $q$, which indicates that  the LDOS difference 
of the electron and hole quasiparticles, shown in Fig.~\ref{fig:DOS_diff},  is not the only factor affecting it. 
However, as one can see by comparing Figs.~\ref{fig:Gevq}(c) and (d),  we find that the total non-local differential conductance 
$G(eV)=\sum_q G(eV,q)$ is mostly  negative (positive) for asymmetric (symmetric) doping. }

\begin{figure*}[t]
    \centering
    \includegraphics[width=0.7\linewidth]{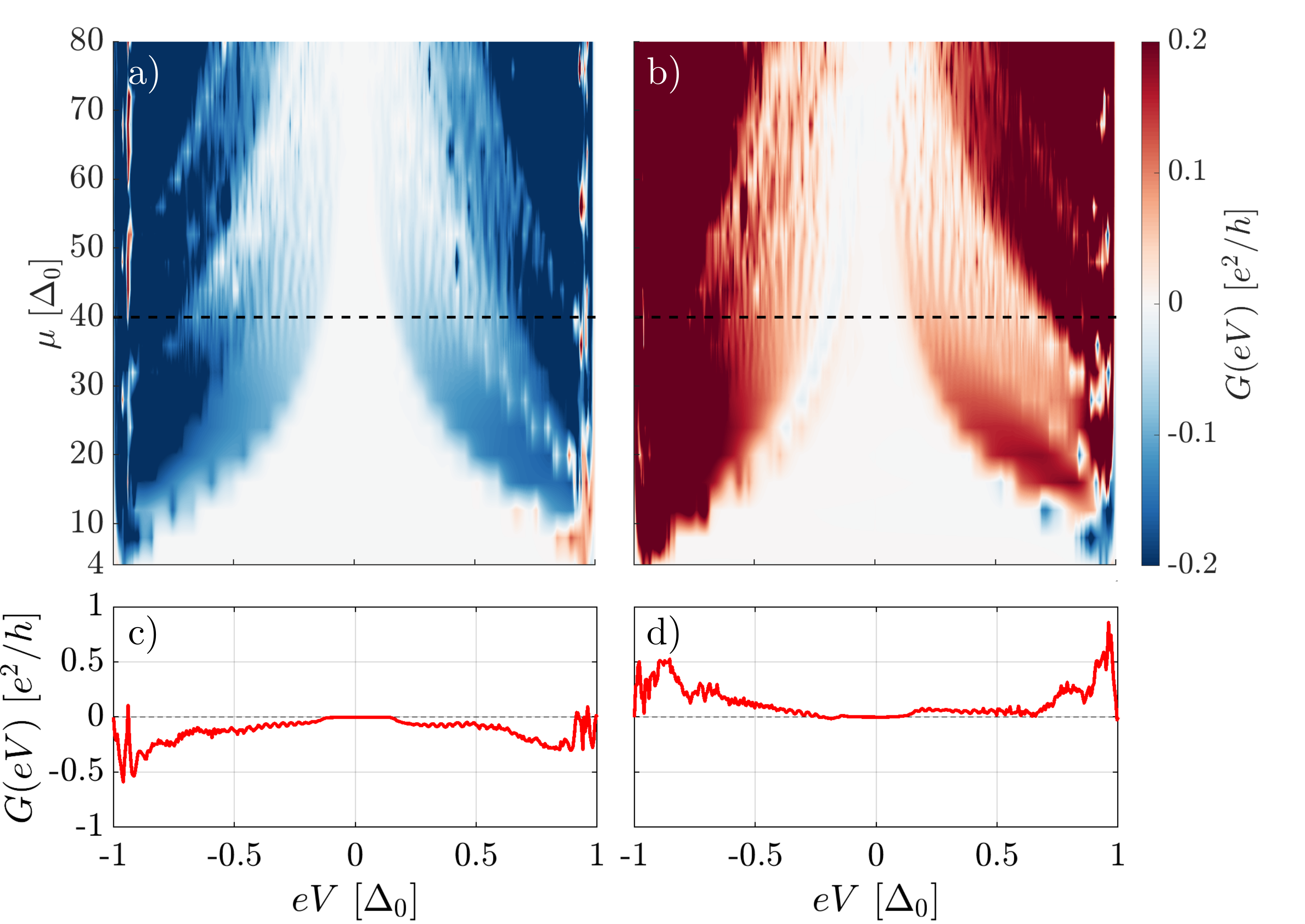}
    \caption{
  The non-local differential conductance $G(eV)$ as a function of the doping of the graphene layers $\mu$ and the applied voltage $V$, at superconducting phase difference $\varphi=0$, and Thouless energy $E_{Th}=3\Delta_0$. In a) an asymmetric doping profile $\mu_b=-\mu_t=\mu$ is used, while in b) the doping is symmetric $\mu_b=\mu_t=-\mu$. c) and d):  $G(eV)$ trace along the dashed line at $\mu=40\Delta_0$ in a) and b), respectively. Negative $G(eV)$ indicates CAR dominated transport.}
    \label{fig:G-mu}
\end{figure*}

\begin{figure}[t]
    \includegraphics[width=\linewidth,trim={0 0 4cm 0},clip]{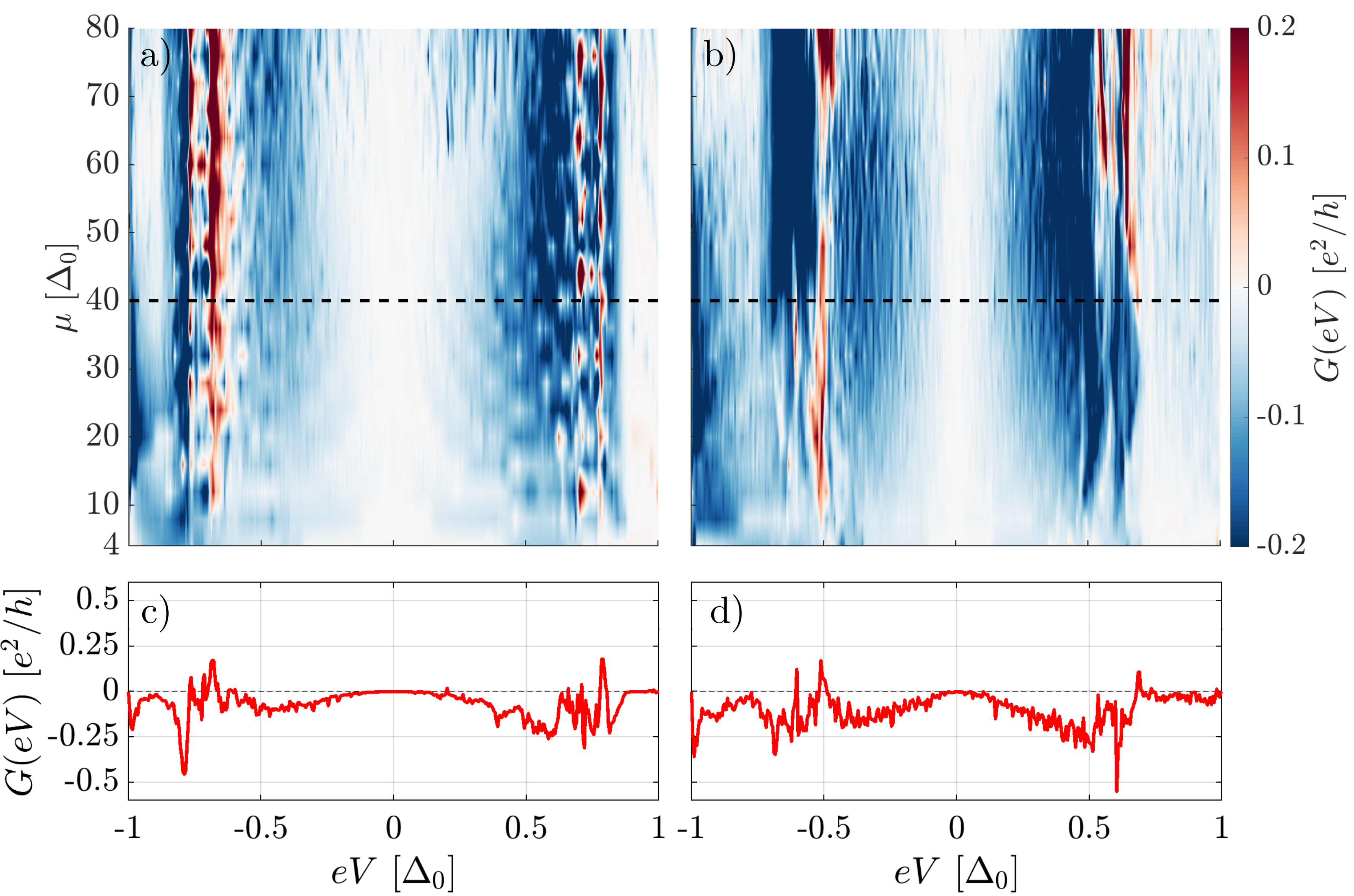}
    \caption{
  The non-local differential conductance $G(eV)$ as a function of the doping of the graphene layers $\mu$ for an asymmetric doping profile in the long junction regime. In a) we used $E_{Th}=\Delta_0$ and in b) $E_{Th}=0.6\Delta_0$ and  the legend for different colors is given in Fig. \ref{fig:G-mu}. The superconducting phase difference is $\varphi=0$. c) and d): $G(eV)$ trace along the dashed line at $\mu=40\Delta_0$ in a) and b), respectively.}
    \label{fig:G-L-mu-big}
\end{figure}

{Next, we study the dependence of $G(eV)$ on the magnitude of the doping of the layers.
In Fig.~\ref{fig:G-mu} we fixed the superconducting phase  difference at $\varphi=0$ and show the results for a setup with a large Thouless energy $E_{Th}=3\Delta_0$. }
The white region around $eV=0$, where $G(eV)$ vanishes,  corresponds to  $|eV|\leq \Delta_{ind}$. 
For low doping, when $\mu\lesssim 4\Delta_0$, the induced gap is almost the same as the bulk gap, i.e., $\Delta_{ind}\approx \Delta_0$ and $G(eV)\approx 0$.
$\Delta_{ind}$ decreases as the doping is increased, and for energies $\Delta_{ind}\leq|eV|\leq\Delta_0$, CAR dominated differential conductance appears 
for the asymmetric doping case (Fig.~\ref{fig:G-mu}(a)). In contrast, as shown in Fig.~\ref{fig:G-mu}(b) for symmetric doping $G(eV)$ is usually positive, 
indicating EC dominated transport.  
We emphasize that contrary  to the  $p$-$n$ junction setup suggested by Ref.~\cite{Cayssol},  
in our setup the doping of the graphene layers does not have to be smaller than $\Delta_0$, which is  experimentally difficult to achieve. 
The CAR dominated transport appears for dopings $\mu > \Delta_0$, when $\Delta_{ind}<\Delta_0$.
We performed similar calculations as in Fig.~\ref{fig:G-mu}(a) for longer junctions as well, see Fig.~\ref{fig:G-L-mu-big}(a) and \ref{fig:G-L-mu-big}(b). 
We find extended regions of CAR dominated transport  when the layers are asymmetrically doped and $\Delta_{ind}<\Delta_0$ is satisfied. 

\begin{figure}[t]
    \includegraphics[width=\linewidth,trim={0 0 4cm 0},clip]{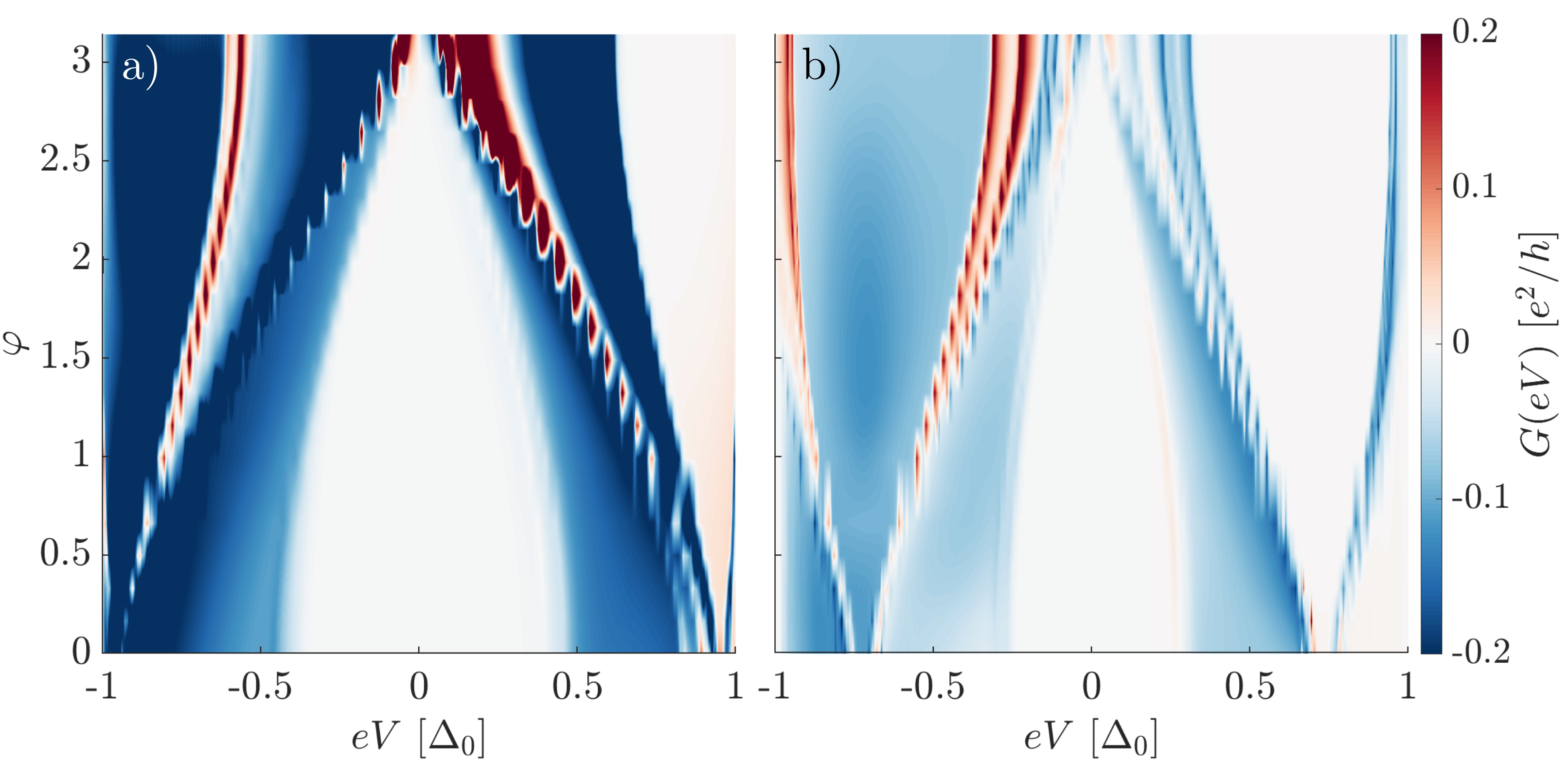}
    \caption{
   The differential conductance $G(eV)$ as a function of the superconducting phase difference $\varphi$, for a) a device in the short junction regime, 
   $E_{Th}=3\Delta_0$ and doping $\mu_b=-\mu_t=20\Delta_0$, and b) $E_{Th}=\Delta_0$ and $\mu_b=-\mu_t=5\Delta_0$.   The legend for different colors is given in Fig. \ref{fig:G-mu}.}
    \label{fig:G-phi}
\end{figure}

As mentioned previously, the superconducting phase difference  $\varphi$ can be another way to tune the non-local transport. 
Typically, the Josephson junction where $\varphi$  should be tuned is part of a large SQUID loop~\cite{Srijit-CPR,PhysRevLett.99.127005}. The magnetic field used in e.g., Ref.~\cite{Srijit-CPR} to change $\varphi$ was of the order of $0.05\,$mT. Such low magnetic fields should have negligible orbital effects 
 in the top and bottom graphene layers, therefore we do not include it explicitly, i.e., through a vector potential $\mathbf{A}(\mathbf{r})$, 
 in the following calculations. We assume that the  only  relevant effect of the magnetic field is  to change $\varphi$ in the Josephson junction.}

{We discuss the  $\varphi$ dependence of the differential conductance} in the calculations shown in Fig.~\ref{fig:G-phi}(a), where we used the same $E_{Th}$ as in Fig.~\ref{fig:G-mu}(a), whereas  Fig.~\ref{fig:G-phi}(b)
corresponds to the case in Fig.~\ref{fig:G-L-mu-big}(a).
We remind that as $\varphi$ increases from $0$ to $\pi$, the induced gap in the graphene layers  is gradually reduced and $\Delta_{ind}$ goes to zero for $\varphi=\pi$, see Figs.~\ref{fig:DOS-1L-2L}(d)-(f).
This appears as a shrinking, low-conductance white region for $|eV|\lesssim\Delta_{ind}$ in Fig.~\ref{fig:G-phi}(a) and \ref{fig:G-phi}(b). 
However, $G(eV)$ is finite and negative 
in the range $\Delta_{ind}\leq|eV|\leq\Delta_0$ for most values of $\varphi$, suggesting that CAR is also robust to the change of $\varphi$. 
Similar behavior can be seen for both $E_{Th}=3\Delta_0$ and $E_{Th}=\Delta_0$.
We have checked that for symmetric doping $\mu_t=\mu_b$  the differential conductance is mostly positive for all values of $\varphi$, i.e., the interlayer transport is dominated by EC. 

\section{Conclusion}
\label{sec:conclusion}

In conclusion, we have studied non-local Andreev reflection in a monolayer graphene based double JJ geometry. 
We have shown, that the ABSs appearing in the graphene layers hybridize and form AMSs. By studying the non-local differential conductance, we found that choosing an 
asymmetric doping profile in the graphene layers leads to CAR dominated transport mediated by the AMSs. 
Changing the doping profile to a symmetric one leads to the suppression of CAR. Importantly, the observed  negative differential conductance does not 
require a very low doping of the graphene layers, which is difficult to achieve. We found that the negative non-local differential conduction  is robust with respect  
to the junction length, changes in the doping of the graphene layers and the superconducting phase difference. 


\section{Acknowledgments}
\label{sec:acknowledge}
This work was supported by the ÚNKP-22-5 New National Excellence Program of the Ministry for Innovation and Technology from the source of the National Research, 
Development and Innovation Fund and by the  Hungarian Scientific Research Fund (OTKA) Grant No. K134437.  
A.K. and P. R.  acknowledge support from the Hungarian
Academy of Sciences through the Bólyai János Stipendium
(BO/00603/20/11 and BO/00571/22/11) as well.
The research was supported by the Ministry of Innovation and Technology and the National Research, Development and Innovation Office within the
Quantum Information National Laboratory of Hungary and we acknowledge the computational resources provided by the Wigner Scientific Computational Laboratory (WSCLAB). 

\bibliography{bibliography}

\end{document}



\section{Boundary conditions}
\label{sec:further-details}

In our calculations we consider a device that is translation invariant in the $y$ direction. Thus,  the transverse momentum $q$ is a good quantum number. 
{For wide junctions and high dopings, when there are many transverse momenta, the exact form of the boundary conditions does not affect the results
and therefore, we used the infinite mass boundary condition~\cite{PhysRevLett.96.246802,PhysRevB.74.041401}
to obtain $q_n = (n + 1/2 ){\pi}/{w}$, where $n = 0, 1, 2, \dots$ and $w$ is the width
of the junction in the $y$ direction. Thus, the differential conductance in Eq. (4) in the main text can be calculated as a sum over $q_n$ values.
We use $w\approx 8\,\mu$m in our calculations.
}

\section{The superconducting leads and the coherence length $\xi_{\perp}$ }
\label{sec:app_super}

In order to give a clearer picture of the structure of the device, we mention that top and bottom normal graphene layer is attached to 
the topmost and lowermost layer of the superconducting leads $S_L$ and $S_R$. These two layers in the leads were separated by $8$ graphene 
layers, making $S_L$ and $S_R$ $10$ layers thick in total. This corresponds to a vertical distance of $d=3.3$ nm between the two normal graphene layers.
In our numerical calculations the Fermi energy is set to $E_F=0.8$\,eV in $S_L$ and $S_R$, which means that there are many open channels, 
and $\Delta_0=1$\,meV is used for the amplitude of the pair-potential. 

\begin{figure}[htb]
    \renewcommand*{\thefigure}{S\arabic{figure}}
    \centering
    \includegraphics[width=\linewidth]{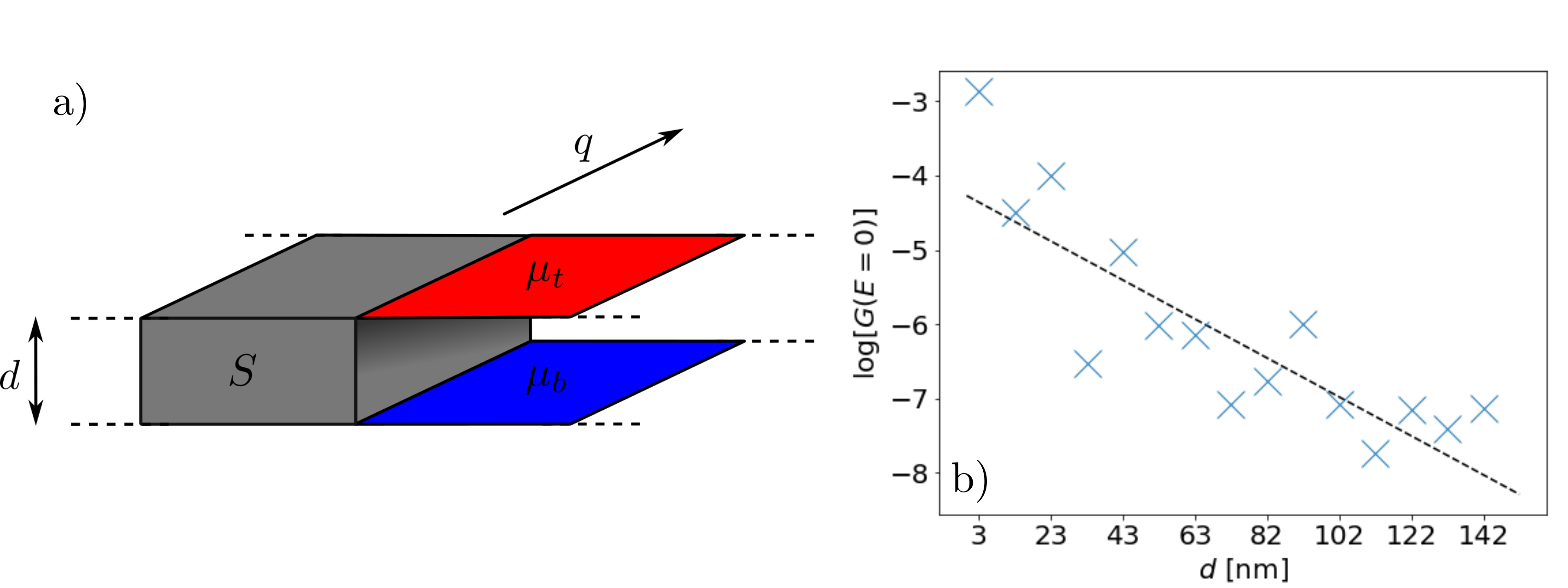}
    \caption{a) The setup used to calculate the interlayer conductance $G_{}(E=0)$ for different thicknesses $d$ of the superconductor. 
    b) The interlayer conductance $G_{}(E=0)$ for different thicknesses $d$ of the superconductor and an exponential fit, giving $\xi_{\perp}\approx38$ nm.}
    \label{fig:interlayer-AR}      
\end{figure}

As mentioned in the main text,  $S_L$ and  $S_R$ were modeled with Bernal stacked graphene, using a minimal tight-binding model, where only the $\gamma_1$ 
interlayer hopping is considered. One can expect that the ABSs formed in the two graphene layers can hybridize if  $d$ is smaller than the coherence 
length $\xi_{\perp}$ in the superconductor.
Therefore, we needed to estimate $\xi_{\perp}$. 
To this end we used a similar approach to Ref.~\cite{Rosdahl2018}. In the  setup shown in Fig.~\ref{fig:interlayer-AR}(a), 
an electron incident from the top graphene layer on the superconductor can be either Andreev reflected into the same layer 
or it can propagate as an evanescent wave $\propto \mathrm{exp}(-d/\xi_{\perp})$ in the superconductor, and reach the bottom graphene layer.  
Here $d$ is the thickness of the superconductor and $\xi_{\perp}$ the coherence length. 
Following Ref.~\cite{Rosdahl2018},  we extract $\xi_{\perp}$ from the decay of the interlayer conductance $G_{}(E=0)$ as $d$ is increased. 
The data is shown in Fig. \ref{fig:interlayer-AR}(b). A coherence length of $\xi\approx38$ nm was obtained, which is greater than the 
thickness of the superconducting lead used in the calculations ($d=3.3$ nm) and subsequently, than the distance between the two normal 
graphene layers connected to it. This is the requirement of coherent CAR. We also found that $\xi_{\perp}$ is independent of the doping of the normal regions.



\section{Derivation of the differential conductance}
\label{sec:app_diffcond}

In this section, following Ref.~\cite{Plaszko2020}, we derive Eq.~(5) from Eq.~(3) of the main text. For simplicity we repeat Eq. (3) here
\begin{equation}
    G(eV,q) = -\frac{2e}{h}\mathrm{Re}\bigg\{\frac{d}{dV}\int dE \, \mathrm{Tr}\big[\tau_z W_{1}  \mathrm{G}^<_{C,1}(E,eV,q)\big]\bigg\}.
    \label{eq:dIdV2}
\end{equation}
To lighten the notations,  the $q$ wavenumber dependence is not shown explicitly in the following. 
We start by inserting the Keldysh equation for the bottom lead--central region lesser Green's function
\begin{equation}
    \mathrm{G}^<_{C,1}(E,eV)=\big[\mathrm{G}^{R}(E)\Sigma^<(E,eV)\mathrm{G}^{A}(E)\big]_{C,1},
    \label{app:G_less}
\end{equation}
into Eq. (3) of the main text, where $\mathrm{G}^{R[A]}(E)$ is the retarded [advanced] Green’s function of the whole device 
and $\Sigma^<(E,eV)$ is the lesser self-energy. 
The differentiation with respect to $V$ in Eq.~\eqref{eq:dIdV2} acts only on the lesser self-energy, which is given by  
$\Sigma^<(E,eV)=\Sigma^<_{S_L}(E)+\Sigma^<_{S_R}(E)+\Sigma^<_1(E)+\Sigma^<_2(E,eV)$. Therefore,  only the lesser self-energy of $N_2$ needs  
to be considered and Eq.~\eqref{eq:dIdV2} takes the form
\begin{equation}
    G(eV) = -\frac{2e}{h}\mathrm{Re}\bigg\{\int dE \;\mathrm{Tr}\bigg[ \tau_3 W_{1} 
     \Big[\mathrm{G}^{R}(E)\frac{d}{dV}\Sigma^<_2(E,eV)\mathrm{G}^{A}(E)\Big]_{C,1} \bigg]\bigg\},
\end{equation}
where the inner matrix product yields
\begin{equation}
    G(eV) = -\frac{2e}{h}\mathrm{Re}\bigg\{\int dE \;\mathrm{Tr}\bigg[ \tau_3 W_{1} 
    \mathrm{G}^{R}_{C,2}(E)\frac{d}{dV}\Sigma^<_2(E,eV)\mathrm{G}^{A}_{t,1}(E) \bigg]\bigg\},
    \label{app:G1}    
\end{equation}
with $\mathrm{G}^{R}_{C,2}(E)$ the top lead--central region retarded Green’s function, and $\mathrm{G}^{A}_{2,1}(E)$ 
the bottom lead--top lead advanced Green’s function. The Green's functions can be further expanded \cite{Claughton1995} as
\begin{equation}
    \mathrm{G}^{R}_{C,2}=\mathrm{G}^{R}_{C,C}W^\dagger_2 \mathrm{g}_2^R,  
    \label{app:greens1}
\end{equation}
\begin{equation}
    \mathrm{G}^{A}_{2,1}=\mathrm{g}_2^A W_2 \mathrm{G}^{A}_{C,C}W^\dagger_1 \mathrm{g}_1^A,
    \label{app:greens2}
\end{equation}
where $\mathrm{G}^{R}_{C,C}$ is the central region retarded Green’s function and $\mathrm{g}_l^{R[A]}$, $l=1,2$ stands 
for the retarded [advanced] Green's function of the normal leads decoupled from the central region. 
The lesser self-energy $\Sigma^<_2(E,eV)$, can be written as
\begin{eqnarray}
    \Sigma^<_2(E,eV)&=&
    \begin{pmatrix}
    f_e(\Sigma_{2,e}^A-\Sigma_{2,e}^R) & 0 \\
    0 & f_h(\Sigma_{2,h}^A-\Sigma_{2,h}^R)
    \end{pmatrix}
    \nonumber\\
    \Sigma^<_2(E,eV)&=&
    \begin{pmatrix}
    f_e\Big((\mathrm{g}_{2,e}^R(E))^{-1} - (\mathrm{g}_{2,e}^A(E))^{-1}\Big) & 0 \\
    0 & f_h\Big( (\mathrm{g}_{2,h}^R(E))^{-1} - (\mathrm{g}_{2,h}^A(E))^{-1} \Big)
    \end{pmatrix},\nonumber
\end{eqnarray}
\vspace{-2cm}
\begin{equation}
    \label{app:eq_self2}
\end{equation}
where $f_e=f(E-eV)$ [$f_h=f(E+eV)$] is the thermal occupation number for the electrons [holes] given by the 
Fermi-distribution function, and $\Sigma_{2,e}^R$ [$\Sigma_{2,e}^A$] and $\Sigma_{2,h}^R$ [$\Sigma_{2,h}^A$] are the retarded 
[advanced] self energies of the electron-like and hole-like particles in the top lead, uncoupled from the rest of the system. 
To calculate the self energies and the Green’s functions we followed the numerical procedure described in Ref.~\cite{Rungger2008}. 
Substituting the Green's functions~\eqref{app:greens1} and \eqref{app:greens2} the self-energy~\eqref{app:eq_self2}, 
in Eq.~\eqref{app:G1} and considering the zero temperature limit one finds that $G(eV)=G^e(eV)+G^h(eV)$, where
\begin{eqnarray}
    G^e(eV)
    &=&\frac{2e^2}{h}\mathrm{Re}\bigg\{\mathrm{Tr}\bigg[ \tau_3 W_{1}  \mathrm{G}^{R}_{C,C}W^\dagger_2 \mathrm{g}_2^R
    \begin{pmatrix}
    \Big(\mathrm{g}_{2,e}^R(eV)\Big)^{-1} - \Big(\mathrm{g}_{2,e}^A(eV)\Big)^{-1} & 0 \\ 0 & 0
    \end{pmatrix}
    \mathrm{g}_2^A W_2 \mathrm{G}^{A}_{C,C}W^\dagger_1 \mathrm{g}_1^A \bigg]\bigg\}
    \label{app:G3}\nonumber\\
    &=&-\frac{4e^2}{h}\mathrm{Im}\bigg\{\mathrm{Tr}\bigg[ \tau_3 W_{1}  \mathrm{G}^{R}_{C,C}W^\dagger_2
    \begin{pmatrix}
    \mathrm{Im}\big(\mathrm{g}_{2,e}^R\big) & 0 \\ 0 & 0
    \end{pmatrix}
    W_2 \mathrm{G}^{A}_{C,C}W^\dagger_1 \mathrm{g}_1^A \bigg]\bigg\}.
    \nonumber
\end{eqnarray}
\vspace{-2cm}
 \begin{equation}
     \label{app:G2} 
 \end{equation}
All energy dependent quantities in Eq. \eqref{app:G2} are evaluated at energy $eV$. 
The expression for $G^h(eV)$ is very similar and therefore not given explicitly. 

Up to this point the derivation is general. We now assume that there is only one AMS in the central region. 
In the presence of the normal leads, the AMS starts to leak out via the normal leads resulting in the broadening of the AMS energy levels. 
Since our main interest are the transport properties below the superconducting gap, in the relevant energy regime we do not expect 
any further bound states in $\mathrm{G}^{R}_{C,C}$ besides the ones corresponding to the AMS. 
We denote by $|AMS\rangle$ the wave function of the AMS, $E_{AMS}$ is the energy of the AMS and define the level broadening 
\begin{equation}
 \Gamma_l^\alpha=-\langle AMS | W_l^\dagger\mathrm{Im}[\mathrm{g}_{l,\alpha}^R(E_{AMS})]W_l | AMS \rangle,
 \label{eq:gamma}
 \end{equation}
 where $\alpha=\{e,h\}$.
Then the Green's function $\mathrm{G}^{R}_{C,C}$ can be approximated by 
\begin{equation}
    \mathrm{G}^{R}_{C,C}(eV)\approx\frac{|AMS\rangle\langle AMS|}{eV-E_{AMS}+\mathrm{i}\Gamma},
    \label{app:approx}
\end{equation}
where $\Gamma=\Gamma_1^e+\Gamma_1^h+\Gamma_2^e+\Gamma_2^h$ is the level broadening of the AMS.
Using Eq.~\eqref{app:G2} one finds 
\begin{eqnarray}
    G^e(eV) &=& -\frac{4e^2}{h}\mathrm{Im}\bigg\{\dfrac{\langle AMS|W^\dagger_2
    \mathrm{Im}\big(\mathrm{g}_{2,e}^R\big)
     W_2| AMS\rangle}{(eV-E_{AMS}+\mathrm{i}\Gamma )(eV-E_{AMS}-\mathrm{i}\Gamma )} \langle AMS|W^\dagger_1 \mathrm{g}_1^A  \tau_3 W_{1}| AMS\rangle\bigg\}\nonumber\\
     &=& -\frac{4e^2}{h}\dfrac{-\Gamma_2^e\,(\Gamma_1^e -\Gamma_1^h)  }{(eV-E_{AMS})^2+\Gamma^2},\nonumber
\end{eqnarray}
\vspace{-2.3cm}
\begin{equation}
    \label{app:Ge} 
\end{equation}
and we also neglect the energy dependence of the Green's function of the normal leads in a $\Gamma$ wide vicinity of the energy $E_{AMS}$ 
in Eq.~\eqref{app:Ge}. The final equation for the differential conductance is obtained by calculating the contribution $G^{h}(eV)$ 
coming from the hole-like degrees of freedom of the self-energy 
in Eq.~\eqref{app:eq_self2}. The total differential conductance thus reads
\begin{equation}
    G(eV)\approx \frac{4e^2}{h}\dfrac{(\Gamma_1^e - \Gamma_1^h )(\Gamma_2^e -\Gamma_2^h)  }{(eV-E_{AMS})^2+\Gamma^2}.
    \label{app:Gev_simple}     
\end{equation}

The calculation can be easily generalized to the case when there are multiple AMSs in the junction. In this case 
\begin{equation}
    \mathrm{G}^{R}_{C,C}(eV)\approx\sum_n\frac{|AMS,n\rangle\langle AMS,n|}{eV-E_{n}+\mathrm{i}\Gamma_n},
\end{equation}
where $E_p$ denotes the energy of the $p$th AMS. The differential conductance reads 
\begin{equation}
    G(eV)= 
    \frac{4e^2}{h}\sum_{n, m}\dfrac{(\Gamma_{1,n m}^e - \Gamma_{1,n m}^h )(\Gamma_{2, m n}^e -\Gamma_{2,m n}^h)  }{\big(eV-E_{n}+i\Gamma_{n n}\big)\big(eV-E_{m}-i\Gamma_{m m})}.
    \label{app:G_pq}
\end{equation}
The $m=n$ terms are Lorentzian resonances, this is the same type of contribution as  Eq.~(\ref{app:Gev_simple}). Furthermore, by generalizing Eq.~(\ref{eq:gamma}) to the 
case of multiple AMSs one can see that $\Gamma_{l, n m}^{\alpha}$, where $n\neq m$,  depends on the interference of the wavefunctions of different AMSs. We remind 
that the energies $E_{n}$, $E_{m}$ and the broadening $\Gamma_{l, n m}^{\alpha}$ depend on the wavenumber $q$.

\bibliography{bibliography}